\documentclass{article}

\usepackage[preprint]{neurips_2023}

\usepackage[utf8]{inputenc} % allow utf-8 input
\usepackage[T1]{fontenc}    % use 8-bit T1 fonts
\usepackage{hyperref}       % hyperlinks
\usepackage{url}            % simple URL typesetting
\usepackage{booktabs}       % professional-quality tables
\usepackage{amsfonts}       % blackboard math symbols
\usepackage{nicefrac}       % compact symbols for 1/2, etc.
\usepackage{microtype}      % microtypography
\usepackage{xcolor}         % colors
\usepackage{graphicx}

\title{Dynamic Grouping for Climate Change Negotiation: Facilitating Cooperation and Balancing Interests through Effective Strategies}

\author{Duo Zhang \\
McGill University \\
845 Rue Sherbrooke O, Montréal, QC H3A 0G4, Canada \\
\texttt{duo.zhang4@mail.mcgill.ca}
\And
Yuren Pang \\
University of Washington \\
Seattle, WA, USA \\
\texttt{ypang2@uw.edu}
\AND
Yu Qin \\
Eccles School of Business \\
University of Utah \\
Salt Lake City, UT, U.S.A. \\
\texttt{yu.qin@eccles.utah.edu}  \\
}

\begin{document}

\maketitle

\begin{abstract}
The current framework for climate change negotiation models presents several limitations that warrant further research and development. In this track, we discuss mainly two key areas for improvement, focusing on the geographical impacts and utility framework. In the aspects of geographical impacts, We explore five critical aspects: (1) the shift from local to global impact, (2) variability in climate change effects across regions, (3) heterogeneity in geographical location and political structures, and (4) collaborations between adjacent nations, (5) the importance of including historical and cultural factors influencing climate negotiations. Furthermore, we emphasize the need to refine the utility and rewards framework to reduce the homogeneity and the level of overestimating the climate mitigation by integrating the positive effects of saving rates into the reward function and heterogeneity among all regions. By addressing these limitations, we hope to enhance the accuracy and effectiveness of climate change negotiation models, enabling policymakers and stakeholders to devise targeted and appropriate strategies to tackle climate change at both regional and global levels.  
\end{abstract}

\section{Consideration of Geographical Impacts}
An existing drawback of the \texttt{RICE-N} framework is the failure to consider the geographical and topological interrelations. This essay will scrutinize four crucial aspects that could remedy this limitation: 
(1) the shift from local to global impact; (2) the variability in the effects of climate change across diverse regions; (3) the heterogeneity in geographical location and political structures; (4) the collaborations between adjacent nations; (5) incorporating historical and cultural factors 

A crucial aspect in the settings of climate change negotiation models that climate change is a chronic process is the chronic transition from local to global consequences. The assumption of homogeneous climate-related features across the world neglects the fact that climate change impacts are often experienced locally before they scale up to the global level. Incorporating this feature into the model would substantially influence the decision-making process of agents during negotiations or proposal-making. For instance, considering local consequences of climate change, such as droughts in sub-Saharan Africa~\cite{IMF_Africa_Food_Insecurity_2022} or flooding in South Asia~\cite{IFRC_Floods_South_Asia_2021}, would enable the model to more accurately represent the priorities and concerns of various countries during negotiations. This feature would enable countries to make heterogeneous short-sighted and long-sighted decisions in climate negotiations. Failure to account for this feature may result in negative impacts on the model's overall performance and the ultimate outcomes of climate negotiations.

The second limitation of current negotiation models is the inadequate consideration of the differential consequences of climate change in different regions. Climate change has divergent consequences for distinct regions and countries, necessitating a regional perspective to develop accurate and effective models. For example, tropical, coastal, or island regions may experience a rising sea level, while colder regions may suffer from thawing permafrost or intensified storms. Integrating these distinct impacts into the model would enable a more comprehensive understanding of the stakes for each country during climate negotiations. A more nuanced approach to modeling the impacts of climate change would not only improve the representation of different countries' priorities but also foster more informed decision-making and potentially lead to more equitable outcomes.

Thirdly, the impact of climate change in various geographic contexts may lead to distinct decision-making frameworks among agents in the model. This implies that not only will the outcomes of these frameworks differ, but also the actions taken by governments in different states. These discrepancies can be attributed to varying political structures and commitments to addressing climate change. However, in the current model, agents use the same utility functions to make decisions despite having multiple parameters that represent their differences. Consequently, the heterogeneity in geographical location and political structures among different regions cannot be entirely captured. For instance, countries with federal political systems, such as the United States and Canada, may approach climate change negotiations differently than countries like China and Russia due to their shared decision-making processes and collaborative goals. Incorporating these distinctions in the \texttt{RICE-N} model would capture the nuances of climate change negotiations among different countries, ultimately improving the overall accuracy and effectiveness of such models.

Fourthly, the current model fails to account for geographical relationship in a region. Climate negotiation often occur in regions; neighboring states are more likely to reach detailed deals that align with regional interests. While the current model includes some form of aggregated economic factors and continents, it fails to consider different geographic collaboration in the reinforcement learning training process, likely leading to a less comprehensive understanding of the climate negotiation process. For example, collaborations between Scandinavian countries in renewable energy and the collective efforts of Pacific Island nations in combating sea-level rise are critical aspects that should be considered in the models to represent the true nature of global climate negotiations. Therefore, it is essential to incorporate geographical relationships to gain a more comprehensive understanding of regional cooperation in addressing climate change.

Finally, in addition to these four main points, another aspect to consider is the importance of incorporating historical and cultural factors that may influence climate change negotiations. For instance, the historical responsibility of industrialized nations in contributing to climate change may play a significant role in shaping their priorities and commitments during negotiations. Furthermore, cultural factors, such as societal values and beliefs surrounding climate change, may also impact a country's willingness to engage in climate negotiations or adopt certain mitigation measures~\cite{adger2013cultural}. For example, place attachment is an important concept, yet perceived different among cultures. Individuals with a strong attachment to their community are often unwilling to migrate because they are reluctant to leave behind their social and emotional support groups and adapt to a new community~\cite{field1988rural}. Accounting for cultural values (e.g., collectivist countries compared to individualist countries~\cite{hofstede2003cultural}) may lead to different and localized policy change. Including these factors in the model would help to create a more comprehensive representation of the complex interplay of factors shaping climate change negotiations.

In conclusion, considering the unique local and regional difference on climate change negotiation is essential for developing more accurate and effective climate change negotiation models. Addressing these limitations would not only enable the models to better represent the complexities of climate negotiations but also facilitate a more realistic understanding of the negotiation process. This, in turn, would allow policymakers and stakeholders to make informed decisions and devise targeted and appropriate strategies to tackle climate change at both the regional and global levels. Ultimately, by incorporating these features, climate change negotiation models can serve as essential tools in guiding international, and regional, efforts to mitigate the devastating impacts of climate change and to promote more equitable, sustainable, and effective climate action.

\section{Improvement of the Utility and Rewards Framework}
The design of utility functions is another aspect that can be improved in the further research. According to the white paper of this project, the reward \(r_{i,t}\) or utility \(U_{i,t}\) of agents are defined by the aggregated consumption \(C_{i,t}\) and the local population \(L_{i,t}\) as shown in equation:
\[r_{i,t} = U_{i,t} = \frac{1}{1 - \alpha} L_{i,t}((\frac{C_{i,t}}{L_{i,t}})^{1- \alpha} - 1)\]
In this equation, the subscripts represent the region \(i\) at the step \(t\), and we can find that the reward primarily depends on the two factors: population and consumption. While population is a local attribute of the regions, the reward is predominantly reliable on the aggregated consumption variable. Furthermore, the aggregated consumption is the cumulative results of domestic consumption and tariff-ed imports. For the domestic consumption \(C_{i,i,t})\), it is calculated as follows:
\[C_{i,i,t} = (1 - s_{i,t})Q_{i,t} - exports\]
In this equation, \(s_{i,t}\) denotes the saving rate, and \(Q_{i,t}\) represents the gross output. Hence, \(s_{i,t})\) primarily exerts a negative impact on the consumption, leading to a further negative effects on the reward function. Additionally, in the term of exports, high saving rates may result in a high value of max potential exports, which also negatively affects on consumption.

As a result, the saving rate in the basic structure of \texttt{RICE-N} only demonstrates negative impacts on the rewards, and this is also the reason why we have a relatively smaller rewards compared with the baseline scenario. Nonetheless, this structure overlooks the positive effects of saving rates on augmenting capital and fostering economic development. Therefore, future developments or research should focus on improving the model structure by incorporating the positive effects of saving rates into the reward function, thereby rendering agents' decision-making processes more comprehensive. This enhancement would allow the model to better capture the nuances of economic interactions and provide a more accurate representation of agents' behavior in the context of resource allocation and consumption.

Another inherent limitation of the \texttt{RICE-N} framework is the assumption of the uniform minimum rates, which denotes the fraction of mitigation efforts undertaken by a country. In \texttt{RICE-N}, there are 27 individual country units, each of which is modeled as an independent decision-making agent. The negotiated agreements governs the minimum mitigation rate for each country. However, the current framework yields a uniform minimum mitigation rate (90\%) for all countries involved, after the \texttt{RICE-N} reinforcement learning process, disregarding regional differences. The current uniformity in the minimum mitigation rate oversimplifies the geographical differences and fails to account for regional variations. For example, requiring the individual countries on the Pacific Islands, such as Tonga and Cook Islands, which have one of lowest $CO_2$ emission to reach a minimum of 90\% would be considerably more arduous than the United States, given their lower base emission quantity. Furthermore, even if the countries with low base emission quantity achieved the minimum mitigation rate, the impact on climate change would likely stay unchanged if more industrial countries such as China and the United States were not committed to such drastic changes in the same time span. As mentioned also in track 2, we think the climate efforts in the baseline scenario are overestimated.   

Our improved dynamic grouping model in Track 2 accounts for the difference in the minimum mitigation rate among different countries after regional grouping.

\begin{table}[h]
\centering
\begin{tabular}{p{2cm}|c|c|c|c|c|c|c|c|c|c|c|c|c|c}
\hline \hline
\centering Region & 1 & 2 & 3 & 4 & 5 & 6 &  7 & 8 & 9 & 10 & 11 & 12 & 13 & 14 \\
\hline
\centering \texttt{RICE-N} & 0.9 & 0.9 & 0.9 & 0.9 & 0.9 & 0.9 & 0.9 & 0.9 & 0.9 & 0.9 & 0.9 & 0.9 & 0.9 & 0.9 \\
\centering \textbf{Dynamic Grouping}  & 0.9 & 0.9 & \textbf{0.6} & \textbf{0.2} & 0.9 & \textbf{0.8} & \textbf{0.7} & \textbf{0.7} & \textbf{0.7} & \textbf{0.5} & 0.9 & \textbf{0.7} & \textbf{0.7} & \textbf{0.7} \\
\hline \hline
\centering Region & 15 & 16 & 17 & 18 & 19 & 20 & 21 & 22 & 23 & 24 & 25 & 26 & 27 \\
\hline
\centering \texttt{RICE-N} & 0.9 & 0.9 & 0.9 & 0.9 & 0.9 & 0.9 & 0.9 & 0.9 & 0.9 & 0.9 & 0.9 & 0.9 &0.9 \\
\centering \textbf{Dynamic Grouping} & \textbf{0.6} & \textbf{0.1} & \textbf{0.7} & \textbf{0.4} & \textbf{0.2} & \textbf{0.7} & 0.9 & \textbf{0.7} & \textbf{0.6} & \textbf{0.6} & \textbf{0.7} & \textbf{0.7} & 0.9 \\
\hline\hline
\end{tabular}
\label{subtable1}
\caption{The minimum mitigation rates for each state using the original \texttt{RICE-N} model and our dynamic grouping model. The minimum mitigation rates are different for certain regions (highlighted), compared to the baseline \texttt{RICE-N} model. }
\end{table}

\section{Conclusion}
In conclusion, we propose two areas for improvement --- consideration of the geographical impact and the improvement of the utility and rewards frameworks. While the \texttt{RICE-N} framework provides a reasonable baseline to simulate climate change negotiation, our proposal adds nuance that considers real-world challenges of climate negotiation among different regions. Our proposal explains the conceptual reasons for considering regional negotiation in lieu of treating individual states as the sole unit of analysis during climate change negotiations. We also introduce concrete implementation recommendation to improve the utility and rewards framework to account for the regional negotiation. We hope that our proposal in this track will lead to more accurate simulations of climate change negotiations.

\bibliography{neurips_2023}
\bibliographystyle{plainnat}

\end{document}